\documentclass[conference]{IEEEtran}
\pdfoutput=1
\IEEEoverridecommandlockouts
\usepackage{amsmath,amssymb,amsfonts}

\usepackage{algorithm}
\usepackage{algorithmicx}
\usepackage{algpseudocode}
\usepackage{graphicx}
\usepackage{textcomp}
\usepackage{xcolor}
\usepackage{epstopdf}
\usepackage[numbers]{natbib}
\usepackage{subfig}
\usepackage{multicol}
\usepackage{lipsum}
\usepackage{enumerate}

\def\BibTeX{{\rm B\kern-.05em{\sc i\kern-.025em b}\kern-.08em
    T\kern-.1667em\lower.7ex\hbox{E}\kern-.125emX}}
\begin{document}

\title{Boosting Byzantine Protocols in Large Sparse Networks with High System Assumption Coverage
\thanks{This work has been accepted by ICPADS2021.\\
\\
© 2022 IEEE.  Personal use of this material is permitted.  Permission from IEEE must be obtained for all other uses, in any current or future media, including reprinting/republishing this material for advertising or promotional purposes, creating new collective works, for resale or redistribution to servers or lists, or reuse of any copyrighted component of this work in other works.}
}

\author{\IEEEauthorblockN{1\textsuperscript{st} Shaolin Yu}
\IEEEauthorblockA{\textit{Tsinghua University}\\
Beijing, China \\
ysl8088@163.com}
\and
\IEEEauthorblockN{2\textsuperscript{nd} Jihong Zhu}
\IEEEauthorblockA{\textit{Tsinghua University}\\
Beijing, China \\
jhzhu@tsinghua.edu.cn}
\and
\IEEEauthorblockN{3\textsuperscript{rd} Jiali Yang}
\IEEEauthorblockA{\textit{Tsinghua University}\\
Beijing, China \\
yangjiali-0411@163.com}
\and
\IEEEauthorblockN{4\textsuperscript{th} Yulong Zhan}
\IEEEauthorblockA{\textit{Tsinghua University}\\
Beijing, China \\
zhanyulong0426@163.com}
}

\maketitle

\begin{abstract}
To improve the overall efficiency and reliability of Byzantine protocols in large sparse networks, we propose a new system assumption for developing multi-scale fault-tolerant systems, with which several kinds of multi-scale Byzantine protocols are developed in large sparse networks with high system assumption coverage.
By extending the traditional Byzantine adversary to the multi-scale adversaries, it is shown that efficient deterministic Byzantine broadcast and Byzantine agreement can be built in logarithmic-degree networks.
Meanwhile, it is shown that the multi-scale adversary can make a finer trade-off between the system assumption coverage and the overall efficiency of the Byzantine protocols, especially when a small portion of the low-layer small-scale protocols are allowed to fail arbitrarily.
With this, efficient Byzantine protocols can be built in large sparse networks with high system reliability.
\end{abstract}

\begin{IEEEkeywords}
Byzantine fault, sparse network, system assumption coverage, system reliability, multi-scale systems
\end{IEEEkeywords}

\section{Introduction}
In real-world distributed systems, the distributed components are often failure-prone.
As it is often hard to show that the undesired failures of these unreliable components would happen with sufficiently low probabilities, these components are often allowed to fail arbitrarily, i.e., being Byzantine \citep{RN2119}, in designing high-reliable fault-tolerant systems.
Meanwhile, by assuming some unit reliability of the distributed components and the independence of component failures in distributed systems, the probabilities of more than some number of the distributed components being simultaneously faulty would be sufficiently low \citep{Powell1992assumption}.
In this background, various kinds of Byzantine-fault-tolerant protocols (\emph{Byzantine protocols} for short) have been proposed in building reliable services with interconnected unreliable components.

However, most of the Byzantine protocols are proposed with the assumption of fully connected networks.
As the numbers of independent communication channels of the distributed components (referred to as the \emph{nodes}) are often practically restricted, these Byzantine protocols should be well extended to networks with low node degrees, especially in large-scale systems \citep{Leighton1992On}.
Unfortunately, as the network connectivity, message complexity, and communication rounds needed for reaching Byzantine-fault-tolerance can hardly be all lowered to satisfy the requirements of real-world applications, the Byzantine protocols are still not widely employed in large networks even with randomization \citep{FM1997,King2011Breaking} and the expense of a portion of \emph{given-up} nonfaulty nodes \citep{Dwork1986,BPG1989,Chandran2010}.

In this paper, to further break the limitations of the Byzantine protocols imposed on node degrees, messages, and time, we try to propose a new paradigm for designing efficient Byzantine protocols in large sparse networks with still high reliability.
Firstly, by identifying the main obstacle in further optimizing the state-of-the-art Byzantine protocols in sparse networks, we propose that the basic assumption of the traditional Byzantine adversaries should be extended in some ways for better evolving in large-scale systems.
Concretely, we would refine the original Byzantine adversary without weakening it.
The adversary can still arbitrarily choose the Byzantine nodes from all the nodes which run the same protocol.
Meanwhile, for constructing multi-scale Byzantine protocols in which several sub-protocols can run in several subsystem scales, a finer \emph{multi-scale adversary} would be defined in better capturing the multi-scale characteristics of large networks.
For this, we would first derive an approximate measurement of the system assumption coverage for constructing the multi-scale adversaries.
Then, by assuming some \emph{sufficiently strong} multi-scale adversaries, we would show that efficient Byzantine protocols can be designed in large sparse networks.
As the system assumption coverage is derived with only the general fault-independence assumption of distributed systems, high reliability can be reached if only this general assumption is not breached in real-world systems.

Comparing with state-of-the-art Byzantine protocols proposed for sparse networks, by adopting some sufficiently strong adversaries, the node degrees, message complexity, and communication rounds of the multi-scale Byzantine agreement are all reduced to logarithmic, which breaks the former limitations on these parameters.
Meanwhile, by refining rather than weakening the adversaries, the results (both including the possibilities and the impossibilities) built upon the classical adversaries are still valid.
With this, classical solutions can be employed as building blocks in playing the game with the finer adversary without losing their tightness in coping with the traditional adversaries.
So, comparing with the \emph{benign} adversary \citep{BIELY20115602}, \emph{random} adversary \citep{BENOR1996329boundeddegree}, and other kinds of weak adversaries, the classical results can be better leveraged with the multi-scale adversaries.
Also, comparing with the randomized solutions \citep{FM1997,King2011Breaking}, the deterministic solutions developed with the multi-scale adversaries can provide a better trade-off between the system assumption coverage and the fault-tolerance efficiency.

The rest of this paper is constructed as follows.
The related work and basic definitions are respectively given in Section~\ref{sec:related} and Section~\ref{sec:model}.
In Section~\ref{sec:obstacle}, the main obstacle of providing Byzantine-fault-tolerance in sparse networks is identified with concrete examples.
With this, an approximate measurement of the system assumption coverage is introduced, and the multi-scale adversary is proposed.
Then, efficient Byzantine protocols are developed with some multi-scale adversaries in Section~\ref{sec:algo}.
Lastly, we conclude the paper in Section~\ref{sec:con}.

\section{Related work}
\label{sec:related}
In the literature, \cite{PSL1980} provides the first Byzantine protocol for reaching deterministic agreement among the unreliable distributed components (or saying the nodes).
From that on, the \emph{Byzantine Generals} problem \citep{RN2119} is widely investigated in synchronous systems with the assumption of a malicious adversary who can arbitrarily choose a portion of the nodes in the system and arbitrarily control these nodes in preventing the other nodes from reaching an agreement.
Generally, it is shown that in tolerating $f$ Byzantine nodes, the number of nodes in the system cannot be less than $3f+1$, the network connectivity cannot be less than $2f+1$, and the deterministic execution time cannot be less than $f+1$ synchronous rounds \citep{Dolev1982StrikeAgain}.
In practice, although these lower-bounds might be acceptable in some small-scale systems, it is hard to apply the classical Byzantine protocols in large-scale systems.

In extending classical Byzantine protocols in large networks, several approaches have been proposed.
Firstly, by \emph{giving up} a small portion of nonfaulty nodes, \emph{incomplete} Byzantine protocols \citep{Dwork1986,BPG1989,Upfal1992,Chandran2010} can be built upon networks with small node degrees.
In this approach, \cite{Dwork1986} shows that deterministic \emph{almost everywhere} Byzantine agreement (BA) can be built upon \emph{bounded-degree} networks with constant node degrees.
\cite{Upfal1992} improves this result with constant Byzantine resilience at the expense of higher computational complexity.
Later in \cite{Chandran2010}, the computational complexity and the \emph{incompleteness} of the secure communication protocols are reduced at the expense of higher node degrees.
However, the scalability of the incomplete Byzantine protocols is still restricted by the overall message complexity, computational complexity, and basic communication rounds.

Secondly, by employing randomization, \emph{randomized} Byzantine protocols \citep{FM1989,FM1997,King2011Breaking} can achieve fast termination or lower message complexity.
In this approach, \cite{FM1989,FM1997} show that the secret-sharing-based \citep{ShareSecret1979} \emph{randomized} BA can terminate in expected constant rounds.
\cite{King2011Breaking} shows that the message complexity of \emph{randomized} BA can be lowered to $o(n^2)$.
However, the required communication rounds and message complexity can hardly be both reduced.
Meanwhile, these protocols are provided for fully connected networks.
Moreover, all randomized protocols are built upon an additional assumption of even distribution and independence of the generated random numbers.
This additional assumption makes the system reliability relying on the realization of the pseudo-random numbers.
On the whole, even with randomization, no Byzantine agreement can reach its goal with sublinear node degree, sublinear message complexity, and sublinear communication rounds at the same time.
These features gravely restrict the applications of higher-layer Byzantine protocols in distributed systems with large numbers of unreliable components.

To further reduce the overall complexity, network connectivity, and communication rounds of fault-tolerant protocols, another approach is to reinvestigate the basic fault assumption.
In \cite{Powell1992assumption}, by establishing a measurement of the \emph{component assumption coverage} for different failure modes, the author argues that the protocols designed with an inappropriate Byzantine fault assumption might be overweighed by protocols designed with some benign fault assumptions.
Thus, instead of handling the overall fault-tolerance problem under the traditional Byzantine adversary, a practical way is to provide the solutions directly with some sufficiently high system reliability.
In this approach, \cite{Steiner2008StartupRecovery} shows that high-reliable hard-real-time systems can be built upon practical Byzantine protocols with restricted failure modes of some communication components.
\cite{Gradient2019} shows that efficient self-stabilizing Byzantine clock synchronization can be built with a restricted Byzantine adversary.
\cite{YuCOTS2021} even shows that an efficient self-stabilizing synchronization solution can be built with standard COTS Ethernet components.
However, all these protocols are built upon some \emph{weak} single-scale adversaries.
With this, the system reliability would depend not only on the algorithms and the unit reliability of the nodes but on the \emph{component assumption coverage} and the restricted power of the single-scale adversary.

\section{Basic model and assumptions}
\label{sec:model}
\subsection{Basic system model}
Generally, the fault-tolerant system $\mathcal{S}$ consists of $n\in \mathbb N$ nodes (denoted as $V$) connected in an undirected network $G=(V,E)$.
In the words of fault-tolerance, each such node can be viewed as a fault-containment region (FCR) in considering the propagation of local faults.
Namely, with the definition of FCR \citep{RN896}, the faults occurring in an FCR cannot be directly propagated to another FCR in the system $\mathcal{S}$.
Nevertheless, the faulty nodes can manifest arbitrary run-time errors as the result of the occurrence of the Byzantine faults and may propagate these \emph{errors} to the nonfaulty nodes or even the whole system, if the protocols running in the system cannot well tolerate the faults occurring in a sufficient portion of the nodes in $\mathcal{S}$.

To design a Byzantine protocol $A$ running in $V$, we assume that the adversary can arbitrarily corrupt a subset $F\subset V$ and make all nodes in $F$ collude together in preventing the nonfaulty nodes $U=V\setminus F$ from reaching their desired goals in $\mathcal{S}$.
Such desired goals can be synchronous agreement, secure communication, reliable broadcast, etc.
Being compatible with \cite{Dwork1986,Upfal1992}, the nonfaulty nodes are also called the correct nodes in the synchronous systems.
By denoting the maximal allowed $|F|$ as $f$, the Byzantine resilience of the protocol $A$ is represented as $\alpha_{A,G}=f/n$.
With classical results \citep{RN2119}, we have $\alpha_{A,G}\in[0,1/3)$.

For simplicity, we assume that the adversary is static and $\mathcal{S}$ is synchronous.
Namely, $F$ is fixed during the execution of $\mathcal{S}$.
Besides, denoting $U=\{1,2,\dots,|U|\}$, the current round state of $\mathcal{S}$ can be represented as $x(k)=(x_1(k),x_2(k),\dots,x_{|U|}(k))$, where $x_i(k)$ is the state of node $i\in U$ in the $k$th round of the execution of $\mathcal{S}$.
Then, by collecting the $k$th round states of all neighbours of $i$ (including $i$), every node $i\in U$ would update its state as $x_i(k+1)$ during the $(k+1)$th round of the execution of $\mathcal{S}$.
In this paper, we only discuss fixed-round executions of $\mathcal{S}$.
With this, the states of $\mathcal{S}$ before the first and after the last rounds of an execution are respectively called the input and output of the execution.

\subsection{Large sparse networks}

Denoting the node degree of each node $i\in V$ in $G$ as $d_i$, $G$ is said to be sparse if $d=\max_{i\in V}\{d_i\}$ is sublinear to $f$.
In other words, we have $\forall{i\in V}:d_i=o(f)$ in sparse networks.
In such networks, as the adversary can corrupt all neighbors of some nonfaulty node $i\in U$ and thus separate $i$ from all other nonfaulty nodes $U\setminus\{i\}$ in the system $\mathcal{S}$, at most a portion of the nonfaulty nodes can reach their desired goal with a fixed Byzantine resilience.
In other words, there would be some nonfaulty nodes being given up in tolerating $f\geqslant d$ Byzantine nodes in the sparse network $G$.
Given the network $G$ and the faulty nodes $F$, the set of all given-up nonfaulty nodes in running the $A$ protocol is denoted as $X_{A}(F,G)$.
Following \cite{Upfal1992}, by denoting $Z_{A}(F,G)=F\cup X_{A}(F,G)$ and $P_{A}(F,G)=V\setminus Z_{A}(F,G)$, it is required that the nodes in $P_{A}(F,G)$ should reach their desired goal in $\mathcal{S}$.
Denoting $x_{A}=\max_{F\subset V}|X_{A}(F,G)|$ with $|F|\leqslant f$, $A$ is said to be an $x_{A}$-incomplete Byzantine protocol in tolerating $f$ Byzantine nodes in $G$ under the traditional adversary.

For a large-scale system $\mathcal{S}$ with hundreds or thousands of nodes, it is common that some protocols only run in a subset of $V$ in $\mathcal{S}$.
In this context, if a protocol $A_0$ runs only in $V_0\subset V$, we assume that no more than $\lfloor \alpha_{A_0,V_0}|V_0|\rfloor$ nodes in $V_0$ can be corrupted by the adversary.
Besides, following the assumption of \emph{independent failure of components} (which is a basic assumption for distributed systems), we assume that the faults that occurred in different nodes of $\mathcal{S}$ are independent with each other.
Following \cite{Powell1992assumption}, by expressing the unit reliability of a node $i\in V$ in some desired duration $\tau>0$ as $r_{i,\tau}=e^{-\lambda_i \tau}$, the failure rate of $i$ during the same duration $\tau$ can be represented as $p_{i,\tau}=1-r_{i,\tau}$.
For simplicity, we assume that all nodes in $V$ share the same unit reliability $r$ during the specific duration $\tau$, and thus the failure rate of every node in $V$ is simplified as $p=1-r$.
In considering practical scenarios, we assume $p\leqslant 10^{-4}$ with $\tau$ being $1$ hour.

\section{The asymmetry and the multi-scale adversary}
\label{sec:obstacle}
In designing Byzantine protocols for large-scale systems, it is crucial to have low complexity, fast termination, affordable networking requirement, low incompleteness, and sufficiently high resilience.
However, these desired properties can hardly be provided simultaneously with the assumption of the traditional adversary.
To ascertain this, an observation of some asymmetry of the sparse networks might be heuristic.
\subsection{The undesired asymmetry}
\label{subsec:asymmetry}
For a concrete example, let us examine the secure communication protocols proposed in the bounded-degree networks.
An interesting observation given in \cite{Upfal1992} shows that the arbitrarily chosen $2t$ faulty nodes cannot contaminate all transmission paths while $t$ such chosen ones can contaminate more than a half of the transmission paths.
Intuitively, this means that the adversary can leverage some asymmetry of the transmission paths.
However, to prevent the adversary from leveraging such asymmetry, we cannot expect to derive some weighted transmission schemes with parallel transmission paths.

To get an intuitive understanding of this, recall that the very initial fault-tolerance problem encountered in bounded-degree networks is that the faulty ones can overwhelmingly surround some correct nodes.
And in the incomplete solutions upon such networks, some correct nodes are allowed to be \emph{poor} (being \emph{given up}) and the remained \emph{non-poor} correct (\emph{npc} for short, also referred to as the \emph{privileged} nodes in \cite{Chandran2010}) nodes are expected to reach their desired goals in the Byzantine protocols.
Does all such \emph{npc} nodes are equivalently \emph{non-poor} in a bounded-degree network?
Obviously, the answer is \emph{no}, since the adversary can place more faulty nodes near some \emph{npc} nodes to make them more \emph{poor} than the other \emph{npc} nodes.

With this intuition, the so-called \emph{non-poor} property might better be extended to some multivalued \emph{luck} property, represented as $\omega(F,i)$ for every node $i\in V$ with the specific $F$.
For example, we can set $\omega(F,i)=0$ if $i\in F$ and define the \emph{npc} nodes as the ones whose \emph{lucks} are beyond some \emph{good-luck} threshold $\omega_0$.
However, as we do not know which nodes would be chosen in $F$ during any concrete execution, we do not know the lucks of the nodes before the execution.
So, for secure communication between two \emph{npc} nodes $i,j$ in playing the game with the traditional adversary, we can only assume that the lucks of $i$ and $j$ being just equal to the threshold $\omega_0$ in considering the \emph{worst-cases}.
Thus, the fact that some pairs of the \emph{npc} nodes might be with better \emph{lucks} than $\omega_0$ cannot be leveraged in designing secure communication protocols.
In this situation, on the one hand, for lower complexity, lower node degrees, and higher resilience, the good-luck threshold $\omega_0$ should be higher.
Nevertheless, on the other hand, for lower incompleteness, $\omega_0$ should be lower.
This dilemma gravely restricts the efficiency of secure communication protocols in large sparse networks.

\subsection{A finer assumption for multi-scale systems}
In offsetting the asymmetry, one possible way is to develop a better \emph{luck} property with a well-balanced \emph{good-luck} threshold in designing specific Byzantine protocols.
However, that would be coupled with the specific goals of the Byzantine protocols.
Alternatively, instead of taking the direction to construct Byzantine protocols only under the traditional adversary, it might make sense to reinvestigate some basic assumptions about the adversary.
Namely, the traditional assumption about the adversary is originally abstracted from fully connected small networks.
In large sparse networks (often with some multiple scales in integrating the building blocks, for example, see \cite{Chandran2010,Jayanti2020}), such assumption seems too coarse to capture the actual properties of the real-world systems.

Concretely, in a large-scale system $\mathcal{S}$, we often want to first construct some small-scale system $\mathcal{S}_0$ with the nodes $V_0$ satisfying $|V_0|\ll n$.
In constructing $\mathcal{S}_0$, we assume only the nodes in $V_0$ being employed.
Now with the assumption of \emph{independent failure of components}, as the failure-rate of every node $i\in V$ is no worse than $p$ for some desired working hours, it would suffice to assume that no more than $\lfloor \alpha_0 |V_0|\rfloor$ faulty nodes (still being arbitrarily chosen by the adversary) with some constant $\alpha_0\in (0,1)$ in satisfying any desired system reliability \citep{Powell1992assumption,Kopetz2004Hypothesis}.
Even when we choose some nodes in $V_0\subset V$ to further construct some other larger-scale systems with $|V|\gg |V_0|$, the assumption of up to $\lfloor \alpha_0|V_0|\rfloor$ faulty nodes in $V_0$ can remain unchanged.
In considering that the added complexity in realizing the nodes in $V_0$ might incur some additional failure-rate in each such node, we can firstly add the worst cases into $ p $.
This makes sense because all qualified real-world devices can provide some constant failure-rate $ p $ despite the various working loads.
So, the innocence of $V_0$ should be defended against the adversary such that, the $V_0$ should pay no more than it deserves in just running any protocol in just the $|V_0|$-scale system.

To be precise, when some protocol only runs with the nodes in $V_0$, for no reason that the adversary can corrupt more than $\lfloor \alpha_0|V_0|\rfloor$ nodes with some constant $\alpha_0$.
Thus, it is better to consider the adversary in some multi-scale context when there are protocols running in more than one scale in the system.
Note that such a \emph{finer} assumption does not contradict the traditional one.
Namely, in the largest scale $n=|V|$, the adversary can still arbitrarily corrupt up to $\alpha n$ nodes in $V$ (the rounding operations are ignored for simplicity when $n$ is large).
Meanwhile, the \emph{multi-scale adversary} can arbitrarily corrupt up to $\alpha_l n_l$ nodes in the protocols running for the $n_l$ nodes.
Generally, the resilience constant can be extended with a resilience function $\alpha:\mathbb N\to \mathbb N$ such that the adversary can arbitrarily corrupt up to $\alpha(s)$ nodes in the given $s$ nodes.

\subsection{A measurement of the system assumption coverage}
So, given the failure-rate $p$ of the unreliable nodes, the critical problem is to provide the resilience function $\alpha$ for the multi-scale system $\mathcal{S}$ with a sufficiently high system assumption coverage.
Here the \emph{system assumption coverage} is extended from the \emph{component assumption coverage} \citep{Powell1992assumption} where the failure modes of the components are the main concern.
Denoting $\mathcal{A}$ as the set of all instances of the Byzantine protocols running in $\mathcal{S}$ and $V_A$ as the set of nodes who run the instance $A\in\mathcal{A}$ in $\mathcal{S}$, the system assumption coverage of $\mathcal{S}$ under $\alpha$ can be represented as
\begin{eqnarray}
\label{eq:assumption coverage} R= \prod_{A\in \mathcal{A}}{Q(\lfloor\alpha(|V_{A}|)|V_{A}|\rfloor,|V_{A}|)}
\end{eqnarray}
where $Q(t,s)$ is a lower-bound of the probability that there are no more than $t$ faulty nodes in the overall $s$ nodes in the distributed system.
With the assumption of \emph{independent failure of components}, $Q(t,s)$ can be generally represented as
\begin{eqnarray}
\label{eq:t_in_s} Q(t,s)= \sum_{i=0}^{t}{\tbinom{s}{i} p^{i}(1-p)^{s-i}}
\end{eqnarray}

With Stirling's approximation $n!\approx \sqrt{2\pi n}(n/e)^n$ \citep{2010Art}, we approximately get
\begin{eqnarray}
\label{eq:approx} \binom{s}{i}\approx \sqrt{\frac{s}{2\pi (s-i)i}}(\frac{s}{s-i})^{s-i}(\frac{s}{i})^i
\end{eqnarray}

Thus, when $s$ is sufficiently large, with $\lim_{s\to \infty}(\frac{s}{s-i})=1$ and $\lim_{s\to \infty}(\frac{s}{s-i})^{s-i}=e^i$, we have
\begin{eqnarray}
\label{eq:approx2} \binom{s}{i}\approx \sqrt{\frac{1}{2\pi i}}(\frac{es}{i})^i
\end{eqnarray}
and thus
\begin{eqnarray}
\label{eq:approx3} Q(t,s)\approx \sum_{i=0}^{t}{\sqrt{\frac{1}{2\pi i}}(\frac{esp}{i})^i (1-p)^{s-i}}
\end{eqnarray}

For the convenience of calculation, as $R$ and $Q(t,s)$ are all very close to $1$, we denote $\nu=1-R$ and $P(t,s)=1-Q(t,s)$.
In our case, as the adversary can arbitrarily choose $F$ with $|F|\leqslant f$ and make all nodes in $F$ fail arbitrarily, $R$ and $\nu$ also respectively represent the system reliability and system failure-rate with respect to the specific working hours.
To calculate $Q(t,s)$, as the ratio of two adjacent items in the right side of (\ref{eq:t_in_s}) can be represented as
\begin{eqnarray}
\label{eq:ratio0} \frac{{\tbinom{s}{i+1} p^{i+1}(1-p)^{s-i-1}}}{{\tbinom{s}{i} p^{i}(1-p)^{s-i}}}= \frac{(s-i)p}{(i+1)(1-p)}
\end{eqnarray}
we have
\begin{eqnarray}
\label{eq:ratio} P(t,s)<\frac{\tbinom{s}{t} p^{t}(1-p)^{s-t}}{\beta-1}\approx {\sqrt{\frac{1}{2\pi t}}(\frac{esp}{t})^t \frac{(1-p)^{s-t}}{\beta-1}}
\end{eqnarray}
when $p\leqslant 1/(\beta s+1)$ holds.
By taking $\beta=2$ and $s<5000$, (\ref{eq:ratio}) would always hold with $p\leqslant 10^{-4}$.
Generally, any larger $s$ can also be handled by summing up the first $\beta ps$ items in calculating $Q(t,s)$.

From (\ref{eq:ratio}) we can see that, with the increase of $t$, $P(t,s)$ soon becomes negligible.
But when $t$ is small, $P(t,s)$ may have a significant effect on the overall system reliability.
So, to develop multi-scale systems, the main difficulty is to provide the small-scale protocols with high resilience.
Given such small-scale protocols, the larger-scale protocols can be built with a much-relaxed resilience function $\alpha$ for the larger $s$.

\section{Solutions and analysis}
\label{sec:algo}
In this section, we give some concrete examples of constructing multi-scale systems with multi-scale adversaries.

\subsection{Immediate Byzantine broadcast}
Firstly, as a simple and practical example, we show that with the assumption of a two-scale adversary, the logarithmic-round deterministic immediate Byzantine broadcast can be reached in logarithmic-degree networks with constant complexity.
Here, when the \emph{General} (correct or faulty) initiates the broadcast, the desired goal is reached iff
1) all correct nodes agree on the same value at the end of the same (finite) round and
2) all correct nodes agree on the value of the correct \emph{General}.

\begin{figure}[htbp]
\centerline{\includegraphics[width=3.3in]{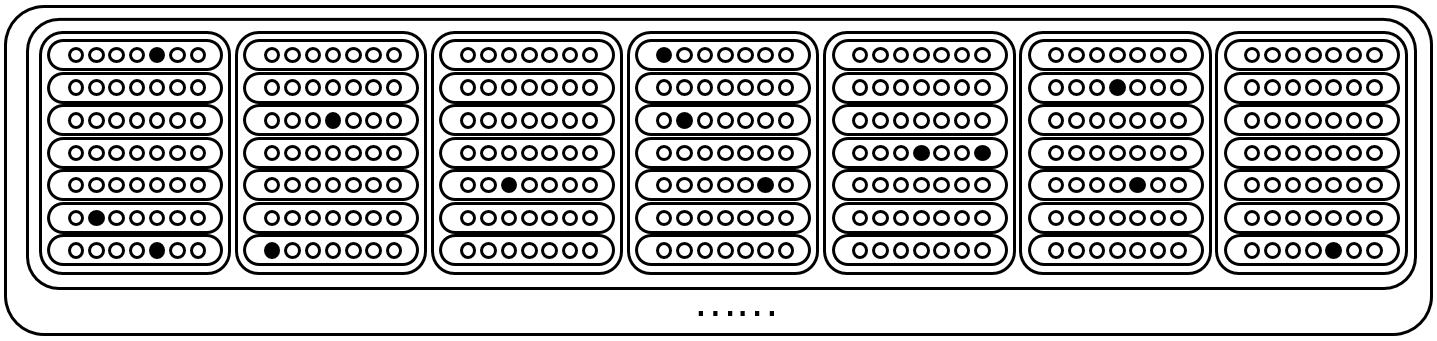}}
\caption{The sparse network $G_\mathtt{H7}$ in the bird's eyes.}
\label{fig:complementary2}
\end{figure}

For this, the sparse network $G=(V,E)$ can be formed as an $s$-base hypercube $G_{\mathtt{H}s}$, as is shown in Fig.~\ref{fig:complementary2}.

In the $s$-base hypercube $G_{\mathtt{H}s}$ with $s=7$, each node is labeled with a $7$-base digital number and represented as a small circle in Fig.~\ref{fig:complementary2}.
Following the basic definition of a hypercube, for any two nodes $i,j\in V$, there is an edge $(i,j)$ on the undirected $G_\mathtt{H7}$ iff the labels of $i$ and $j$ are with one and only one different digit.
For example, in a $3$ dimensional $7$-base hypercube, the node $320$ is connected to the node $321$ and node $620$ but not connected to the node $230$ or node $231$.
As $G_\mathtt{H7}$ has at most $L=O(\log n)$ dimensions, $G_\mathtt{H7}$ is an $O(\log n)$-degree network.
By representing the $k$th dimension position of node $i$ in $G_\mathtt{H7}$ as the $k$th leftmost digit in the label of $i$, the node $a_L\cdots a_2 x$ with $x\in\{0,1,\dots,L-1\}$ form a $7$-node complete graph $K_7$ in the innermost (the first) dimension of $G_\mathtt{H7}$ shown in Fig.~\ref{fig:complementary2}.
As all $7$ nodes in an innermost $K_7$ are labeled with the same rightmost $L-1$ digits, these $L-1$ digits are used to label the innermost $K_7$.
With this, the node $a_L\cdots a_2 x$ is at the $x$ site in the $K_7$ labeled as $a_L\cdots a_2$.
For simplicity, the leftmost $0$ digits in a label can be omitted.

With $G_\mathtt{H7}$, the multi-scale Byzantine broadcast protocol can be constructed as follows.
Firstly, every node $i\in U$ would run one and only one $7$-node BA protocol $A_7$ in the innermost dimension of $G_\mathtt{H7}$ during the execution of $\mathcal{S}$.
By assigning one node $i_0\in V$ as the \emph{General}, the $7$ neighbors of $i_0$ (including $i_0$) in the innermost dimension of $G_\mathtt{H7}$ shown in Fig.~\ref{fig:complementary2} are said to be in the $0$ layer.
In the $0$ layer, the \emph{General} $i_0$ initiate its $0$ layer $7$ neighbors (denoted as $V_0$) with the current state of $i_0$ by running a very simple initiation protocol $I_7$.
Without loss of generality, let us assume the top-leftmost node $0$ in Fig.~\ref{fig:complementary2} being the \emph{General}.
Then, the $0$ layer BA protocol $A_7$ would be performed in $V_0$ (in the top-leftmost $K_7$ labeled with $0$) and would terminate in constant rounds.
At the termination of the $0$ layer BA, each node $j\in U\cap V_0$ would set its state with the agreed value and then initiate the $1$ layer $7$ neighbors of $j$ (in the vertical directions in Fig.~\ref{fig:complementary2}) with the current state of $j$ by running the same $I_7$ protocol.
With this, the nodes (denoted as $V_1$) in the other leftmost $6$ innermost $K_7$ (labeled from $1$ to $6$) would all be initialized.
Then, a differential BA protocol $B_7$ \citep{Garay2003} would be parallel performed in each initialized innermost $K_7$ with constant rounds.
Similarly, at the termination of these differential BA instances, each node $j\in U\cap V_1$ would run the $I_7$ protocol to initiate the $2$ layer $7$ neighbors of $j$ (in the horizontal directions in Fig.~\ref{fig:complementary2}).
With this, the nodes (denoted as $V_2$) in the other $6$ columns (except the ones represented by the ellipsis) would all be initialized.
Then, the differential BA protocol $B_7$ would be parallel performed in each initialized innermost $K_7$ (labeled from $10$ to $66$) with constant rounds.
Iteratively, this procedure would be performed until the $(L-1)$ layer differential BA terminates, with which the agreed value of the BA instances run in every $j\in U$ would be the final output of the overall protocol.
So, the overall protocol can terminate in $O(\log n)$ rounds with $O(1)$ complexity.

Now we show how this protocol can reach Byzantine broadcast.
Firstly, in the $0$ layer, the adversary is allowed to arbitrarily corrupt up to $2$ nodes in the innermost $7$ nodes.
With this, the $0$ layer BA instance can run correctly and output the agreed value for the nodes in $U\cap V_0$.
Then, in running the $I_7$ protocol between every innermost $K_7$ in $V_1$ and the innermost $K_7$ in $V_0$, by denoting the sites in the $K_7$ labeled with $w$ as $S_w$, the adversary is allowed to arbitrarily corrupt up to $2$ sites in $S_{w_1}\cup S_{w_2}$ when $w_1$ and $w_2$ has only one digit being different (or saying $w_1$ and $w_2$ are adjacent).
With this,  at least $5$ correct nodes in every innermost $K_7$ can be initiated with the correct agreed value.
So, by performing the $7$-node differential BA \citep{Garay2003}, all correct nodes would have the correct agreed value in every initiated innermost $K_7$.
Thus, by iteratively applying this result, all correct nodes would have the correct agreed value at the end of the execution of the overall protocol.
Here, for reaching efficient deterministic Byzantine broadcast, a two-scale adversary is defined for the $7$-node BA protocols ($A_7$ and $B_7$) and the $14$-node initiation protocol $I_7$ (for two adjacent innermost $K_7$).
With this, it is easy to see that the overall protocol can be extended to the $s$-base hypercube $G_{\mathtt{H}s}$ under the same two-scale adversary defined for the $s$-node BA protocols and the $2s$-node initiation protocol.

Now we show how this adversary can be supported with practical system assumption coverage.
Firstly, to support the fault-assumption of the $s$-node BA instances, the probability is no less than $(1-P(\lfloor \alpha(s)s\rfloor,s))^{n/s}$ with $\alpha(s)=1/3$.
For the case $s=7$, we have $P(2,7)\approx {\sqrt{\frac{1}{4\pi}}(\frac{7ep}{2})^2 (1-p)^{5}}<40p^2$.
For the larger $s$, we generally have
\begin{eqnarray}
\label{eq:p_1}P(s/3,s)\approx {\sqrt{\frac{3}{2\pi s}}(3ep)^{s/3} (1-p)^{s-s/3}}<(3ep)^{s/3}
\end{eqnarray}

Secondly, to support the fault-assumption of the initiation instances, the probability is no less than $(1-P(\lfloor \alpha(s)s\rfloor,2s))^{n/s-1}$ with $\alpha(s)=1/3$.
For the case $s=7$, we have $P(2,14)\approx {\sqrt{\frac{1}{4\pi}}(7ep)^2 (1-p)^{5}}<160p^2$.
For the larger $s$, we generally have
\begin{eqnarray}
\label{eq:p_2}P(s/3,2s)\approx {\sqrt{\frac{3}{2\pi s}}(6ep)^{s/3} (1-p)^{2s-s/3}}<(6ep)^{s/3}
\end{eqnarray}

So, put it together, we get
\begin{eqnarray}
\label{eq:broadcast_r}R= (1-P(\lfloor s/3\rfloor,s))^{n/s}(1-P(\lfloor s/3\rfloor,2s))^{n/s-1}\nonumber\\
\approx (1-\sqrt{\frac{3}{2\pi s}}((3ep)^{s/3}+(6ep)^{s/3}))^{n/s}\nonumber\\
>(1-(6ep)^{s/3})^{n/s}
\end{eqnarray}

Now, to see how $R$ can be sufficiently high, let us take $s=16$ and $p=10^{-4}$.
In this case, we would have $R\geqslant 1-10^{-9}$ if only $n\leqslant 10^6$.
So, efficient multi-scale Byzantine broadcast protocols can be practically built upon sparse networks with high reliability.

\subsection{Immediate Byzantine agreement}
Given a specific \emph{General}, the Byzantine broadcast protocol provided above performs the immediate reliable broadcast of the \emph{General} in sparse networks with the two-scale adversary.
With this, we show how to build efficient Byzantine agreement in sparse networks with the same adversary.
Here, with every correct node $i\in U$ being initiated with a value $v_i$, the desired goal is reached iff
1) all correct nodes agree on the same value at the end of the same (finite) round and
2) all correct nodes agree on the value $v$ if $\forall i\in U:v_i=v$.

To build Byzantine agreement in the same sparse network $G_{\mathtt{H}s}$, the multi-scale Byzantine broadcast protocol can parallel run for every node $i\in V$ being the \emph{General}.
For efficiency, instead of running $n$ parallel Byzantine broadcast instances for the $n$ nodes, these instances can be run for the $n/s$ innermost $K_s$.
Concretely, in the first round, only $n/s$ $s$-node BA instances would be executed in the $n/s$ innermost $K_s$.
At the end of the first round, by running the $2s$-node initiation protocol $I_s$ for every pair of adjacent innermost $K_7$ in the $1$ layer, the agreed innermost $K_7$ can be viewed as a locally agreed super-node.
Thus, there would be at most $n/s$ Byzantine broadcast instances being parallel run in every correct node of $\mathcal{S}$ during the execution of $\mathcal{S}$.
Then, at the end of the last round, every node $i\in U$ can finally agree on the median of the $n/s$ output values of the $n/s$ Byzantine broadcast instances.

It is easy to see that this protocol reaches the goal of the deterministic immediate Byzantine agreement.
For efficiency, as there are at most $n/s$ Byzantine broadcast instances being run in parallel, the overall complexity would at most be $O(n)$, where the message complexity would be $O(\log n)$, as the messages generated for the parallel instances during the same round in every $O(\log n)$-degree node can be merged into one round-message.
Meanwhile, the required rounds, node-degrees, and system assumption coverage (also system reliability) of the Byzantine agreement protocol are all the same as the provided multi-scale Byzantine broadcast protocol.
So, deterministic $O(\log n)$-round Byzantine agreement can be reached in $O(\log n)$-degree network with $O(\log n)$ message complexity with high reliability.

\subsection{Incomplete Byzantine protocols}
One defect of the multi-scale Byzantine protocols presented above is that the system reliability is built upon the assumption coverage of all employed sub-protocols in all related scales.
In this situation, if the fault-assumption of any employed protocol is breached in any running instance, the overall system may fail.
To avoid this, we show how multi-scale Byzantine protocols can be built with tolerating the failure of some instances of the low-layer protocols.
As an intuitive example, here we investigate the classical secure communication in sparse networks.
For this, by denoting $i,j$ as two \emph{npc} nodes, the desired goal is reached iff
1) there are sufficient \emph{npc} nodes and
2) every message sent from $i$ can be correctly received by $j$ in some finite synchronous rounds and \emph{vice versa}.

\begin{figure}[htbp]
\centerline{\includegraphics[width=2.6in]{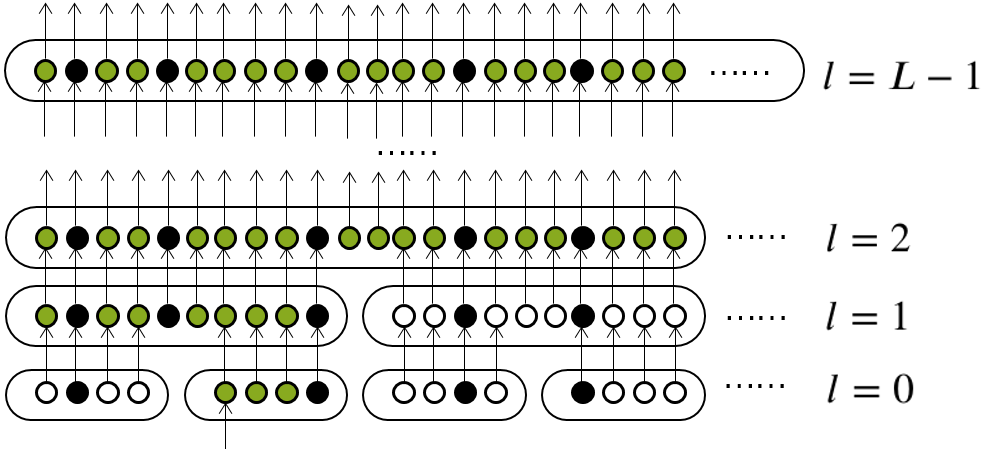}}
\caption{The multi-layer expander $G_{\mathtt{EX}4}$.}
\label{fig:complementary1}
\end{figure}

To construct the overall protocol, we would extend the constant-resilience protocol proposed in \cite{Upfal1992} as the core building block.
Concretely, as the transmission scheme proposed in \cite{Upfal1992} incurs high computational complexity, here we focus on reducing the computational complexity of \cite{Upfal1992}.
For this, the sparse network $G=(V,E)$ can be formed as a multi-layer expander $G_{\mathtt{EX}s_0}$, as is shown in Fig.~\ref{fig:complementary1} with $s_0=4$.

In Fig.~\ref{fig:complementary1}, it should be noted that all the vertical $L$ layers of $G_{\mathtt{EX}4}$ are implemented in just one layer communication nodes.
Namely, the $L$ small circles in every vertical line of Fig.~\ref{fig:complementary1} represent the same communication node.
In other words, the multi-layer expander $G_{\mathtt{EX}s_0}$ is actually a one-layer expander with $n=|V|$ communication nodes, each of which would act as $L$ different logical nodes in running the sub-protocols in the different layers.

At the $0$ layer, there are $r_0= n/s_0 $ independent subnetworks $G_{0,r}=(V_{0,r},E_{0,r})$ with disjoint node-sets $V_{0,r}$ for $1\leqslant r\leqslant r_0$, where $s_0$ is a pre-configured constant.
Each subnetwork $G_{0,r}$ is an $s_0$-node $d_0$-regular expander for running the $0$ layer Byzantine protocols.
For simplicity, we assume that $n$ is divisible by $s_0$ (otherwise, we can make up a slightly larger upper layer and only use the extra upper-layer nodes to run the high-layer protocols, the same below).
Then, at the $1$ layer, there are $r_1= n/s_1 $ independent subnetworks $G_{1,r}$ with $1\leqslant r\leqslant r_1$, where $s_1$ is also a pre-configured constant.
Each subnetwork $G_{1,r}$ is an $s_1$-node $d_1$-regular expander with $s_1=s_0/\theta_1$ ($\theta_1\in (0,1)$) and contains $ 1/\theta_1$ $0$ layer subnetworks.
Iteratively, by configuring the constant $s_{l}=s_{l-1}/\theta_{l}$ ($\theta_{l}\in (0,1)$), $r_{l}= n/s_{l} $ $l$ layer subnetworks $G_{l,r}$ with $1\leqslant r\leqslant r_{l}$ can be formed, each of which contains $ 1/\theta_{l}$ $(l-1)$ layer subnetworks, until it comes the $(L-1)$ layer expander with $s_{L-1}=n$ nodes.
Denoting the state of node $i\in U$ in the $l$ layer as $x_i^{(l)}$, to ensure that each low-layer protocols would not be affected by the upper-layer protocols, the state $x_i^{(l)}$ can only propagate to the state $x_i^{(l+1)}$ while the state propagation from $x_i^{(l+1)}$ to $x_i^{(l)}$ is prohibited.

With this, to deliver a message from $i$ to $j$, an $s_0$-node Byzantine protocol $C_0$ (for example, some Byzantine broadcast protocol or some secure communication protocol) would run in the $0$ layer subnetwork $G_{0,r}$ (containing $i$, the same below) to transmit the message of $i$ to all other nodes in $G_{0,r}$.
As $s_0$ is a constant, these instances would terminate in constant rounds with constant complexity.
With this, if the fault-assumption of the $C_0$ protocol is not breached in $G_{0,r}$, the message of $i$ would be correctly received in all \emph{npc} nodes of $G_{0,r}$ (denoted as $P_{C_0}(F_0,G_{0,r})$ and shown as green in the bottom layer of Fig.~\ref{fig:complementary1}).
Then, the $0$ layer state of every node $i_0\in P_{C_0}(F_0,G_{0,r})$ would be propagated to $x_{i_0}^{(1)}$ in the $1$ layer.

From the $1$ layer on, to reduce the complexity, instead of employing the protocol $C_0$, the logical nodes in the $1$ layer would run a new protocol $C_1$ by replacing the high-complexity operation taken in the original protocol proposed in \cite{Upfal1992} as the majority function (i.e., taking the majority values received from all transmission paths).
For this, all the correct nodes in $V_{0,r}$ would transmit the received message of $i$ to all other nodes in $G_{1,r}$.
So, it needs to show that a sufficient number of nodes $i_1\in V_{1,r}$ would receive the correct message of $i$ in more than a half of all transmission paths from $P_{C_0}(F_0,G_{0,r})$ to $i_1$, as long as the fault-assumption of the $C_1$ protocol is not breached in $G_{1,r}$.
As $G_{1,r}$ is a strong expander, this can be supported with a sufficiently large $\theta_1$.
For a simple example, when $\theta_1\approx 1/2$, the transmission scheme can be simplified as directly sending the message of $i$ to the neighbors in $G_{1,r}$.
With this, assuming the fault-assumption of $C_1$ being not breached, as every \emph{npc} node can have more \emph{npc} neighbours than the faulty and \emph{poor} neighbours in $G_{1,r}$, every \emph{npc} node can receive the correct message of $i$ in one round with applying the majority function.
Iteratively, with $\theta_l\approx 1/2$ for $1\leqslant l\leqslant L-1$, the $l$ layer \emph{npc} nodes can receive the correct message of $i$ in $O(l)$ rounds.
As $L=O(\log n)$, all the $L-1$ layer \emph{npc} nodes can receive the correct message of $i$ in $O(\log n)$ rounds.
Furthermore, a finer investigation of the low bound of $\theta_l$ is also within reach.
As is limited here, we leave this for the interested readers.

To measure the system assumption coverage, if the fault-assumption of $C_0$ and $C_1$ is not allowed to be breached, we can calculate the system reliability as
\begin{eqnarray}
\label{eq:broadcast_r1}\prod_{l=0}^{L-1}(1-P(\lfloor \alpha(s_l)s_l\rfloor,s_l))^{\frac{n}{s_l}}
\end{eqnarray}
In this case, the message of $i$ can be correctly received by $j$ if $i$ is a $0$ layer \emph{npc} node and $j$ is an $L-1$ layer \emph{npc} node.
Thus, if $i$ and $j$ are all \emph{npc} nodes in the $0$ layer and the $L-1$ layer (referred to as the overall \emph{npc} nodes), the goal of secure communication can be reached between such $i$ and $j$.

From (\ref{eq:broadcast_r1}) we can see that, as $s_l$ would become larger with the increase of $l$, the items with larger $l$ would soon become negligible.
So, the system reliability mainly depends on the items with the small $l$.
Now, if no more than $t_l$ instances of the secure communication protocols in the $l$ layer are allowed to fail with sufficiently small $l$, only a small portion of the overall \emph{npc} nodes would be affected.
Meanwhile, the items in (\ref{eq:broadcast_r1}) with the small $l$ can be improved as
\begin{eqnarray}
\label{eq:improve_r}\sum_{t=0}^{t_l}\tbinom{r_l}{t}(1-P(\lfloor \alpha(s_l)s_l\rfloor,s_l))^{r_l-t}P(\lfloor \alpha(s_l)s_l\rfloor,s_l)^t
\end{eqnarray}
With this, we can derive the $l$th item of (\ref{eq:broadcast_r1}) as $1-\nu_l$ with
\begin{eqnarray}
\label{eq:ratio_r} \nu_l<\frac{\tbinom{r_l}{t_l} p_l^{t}(1-p_l)^{r_l-t_l}}{s_l-1}\approx \sqrt{\frac{1}{2\pi t_l}}(\frac{er_l p_l}{t_l})^{t_l} /(s_l-1)
\end{eqnarray}
when $p_l=P(\lfloor \alpha(s_l)s_l\rfloor,s_l)\leqslant 1/(n+1)$ holds.
Thus, $1-\nu_l$ would be improved significantly with $t_l\geqslant 2$.

\subsection{Discussion}
\label{subsec:gainloss}
As we have seen, on one side, the assumption of the multi-scale adversary can place the protocol designers at a much-desired position in deriving easier Byzantine solutions.
Without this multi-scale assumption, the efficiency of the Byzantine solutions would be gravely limited with the identified asymmetry property of the sparse networks.
For example, the computational complexity of the secure communication protocol provided in \cite{Upfal1992} is very high.
The overall complexity of the more efficient protocols provided in \cite{Chandran2010} (also investigated in \cite{Jayanti2020}) is at least polynomial.
The resilience provided in \cite{Dwork1986} is relatively low.
On the other side, the system assumption coverage should be carefully calculated in real-world systems.

Nevertheless, we argue that a practical multi-scale adversary is a good starting point in constructing efficient multi-scale Byzantine protocols.
Firstly, in comparing with probabilistic Byzantine protocols \citep{BENOR1996329boundeddegree}, the probabilistic aspects of the multi-scale systems can be well encapsulated in the multi-scale adversary, with which the deterministic solutions can be decoupled with the calculation of the system assumption coverage.
Secondly, the multi-scale adversary can also provide a finer abstraction for the probabilistic properties in multi-scale distributed systems.
With this, the disadvantage of the asymmetric property of the sparse networks can largely be overcome in multi-scale networks.
Thirdly, even when the \emph{weakest point} of the basic multi-scale assumption is violated, i.e., some lower layer networks are corrupted by more faulty nodes than the ones that can be tolerated, multi-scale protocols can be built with tolerating the failure of some lower layer protocols.
Meanwhile, the original assumption of the single-scale adversary can also be included in the multi-scale ones.
Generally, the finer the multi-scale adversary is given, the better balance between the system assumption coverage and the efficiency of the deterministic Byzantine solutions can be expected in large-scale systems.

\section{Conclusion}
\label{sec:con}
In this paper, we have proposed a new paradigm of developing efficient Byzantine protocols for large sparse networks with high reliability.
Firstly, the undesired asymmetry of sparse networks in building efficient Byzantine protocols with the traditional adversary is identified.
In overcoming this asymmetry, multi-scale Byzantine protocols are proposed with the assumption of the so-called \emph{multi-scale adversary}.
In investigating the reliability of the systems developed with such multi-scale adversaries, an approximate measurement of the system assumption coverage is developed.
Then, it is shown that logarithmic-round deterministic BA can be built upon logarithmic-degree networks with logarithmic message complexity and high system assumption coverage.
It is also shown that the system reliability can be further improved with multi-scale Byzantine protocols that can tolerate the failures of low-layer small-scale protocols.
Meanwhile, with the multi-scale adversaries, the measurement of system assumption coverage and the development of deterministic Byzantine protocols are also decoupled.
With this, finer Byzantine protocols can be further developed for various kinds of large sparse networks.

\bibliographystyle{IEEEtran}
\bibliography{IEEEabrv,BPLSNHSAC}

\end{document}